\documentclass[preprint,pre]{revtex4}%
\usepackage{amsfonts}
\usepackage{amsmath}
\usepackage{amssymb}
\usepackage{graphicx}%
\usepackage{epsfig}
\usepackage{epstopdf}
\usepackage{subfig}
\providecommand{\U}[1]{\protect\rule{.1in}{.1in}}

\begin{document}
\title{Thermodynamic Comparison and the Ideal Glass Transition of A Monatomic Systems Modeled as an
Antiferromagnetic Ising Model on Husimi and Cubic Recursive Lattices of the
Same Coordination Number}
\author{Ran Huang \footnote{Correspondence to: ranhuang@sjtu.edu.cn}}
\affiliation{School of Chemistry and Chemical Engineering, Shanghai Jiao Tong University, Shanghai 200240, China}
\author{P.D. Gujrati \footnote{Principle correspondence to: pdg@uakron.edu}}
\affiliation{The Department of Physics, The University of Akron, Akron, OH\ 44325}

\begin{abstract}
Two kinds of recursive lattices with the same coordination number but
different unit cells (2-D square and 3-D cube) are constructed and the
antiferromagnetic Ising model is solved exactly on them to study the stable and
metastable states. The Ising model with multi-particle interactions is
designed to represent a monatomic system or an alloy. Two solutions of the model exhibit the crystallization of liquid, and the ideal glass transition of supercooled liquid respectively. Based on the solutions, the thermodynamics on both lattices was examined. In particular,
the free energy, energy, and entropy of the ideal glass, supercooled liquid,
crystal, and liquid state of the model on each lattice were calculated and
compared with each other. Interactions between particles farther away than the nearest
neighbor distance are taken into consideration. The two lattices show comparable properties on the transition temperatures and the thermodynamic behaviors, which proves that both of them are practical to describe the regular 3-D case, while the different effects of the unit types are still obvious.

\end{abstract}
\date{\today}
\maketitle

\section{Introduction}

The glass transition in which the amorphous state becomes brittle on cooling
or soft on heating has been studied and investigated for many years. The
situation is, however, still unclear and remains controversial \cite{Kauzmann,glass1,glass2,glass3,glass4}. Numerous
models and methods have been developed to study the glass transition \cite{Debenedetti} with
regards to its thermodynamic or dynamic aspect. A concept drawn from the aspect of thermodynamic is the so-called ``equilibrium glass transition", commonly known as the \emph{ideal glass transition}, which is a hypothetical transition believed to occur in the limit of infinitely
slow cooling rate \cite{Kauzmann}. Among the efforts on studying the ideal glass
transition, one method developed in our group involves the exact thermodynamic
calculation of Ising models on the recursive lattices, such as a Bethe lattice \cite{Bethe} or a Husimi lattice (HL) \cite{Husimi}, and has been demonstrated to be a practical methodology to study the ideal glass transition \cite{pdg_prl_reliable,PDG2,PDG3,PDG4}. 

As the name implies, recursive lattices are constructed recursively to be a fractal structure from the basic units of regular lattices to approximate them, and have the advantage to be exactly solvable \cite{exact,12,13}. Among the many types of recursive lattices, HL was developed by K. Husimi and has been employed in various field as a successful physics model. The original form of HL is a fractal recursion of square units connected to each other on the vertex. For several decades it has been extended to be in various forms constructed from different kinds of units, such as triangle, hexagon, tetrahedron, cube, and so on \cite{unit1,Geertsma,unit2}. In this paper, we construct a multi-branched square HL \cite{Ran1} and compare its behavior with a Husimi cubic recursive lattice (CRL), both of which are designed to approximate the regular 3-D cubic lattice whose coordination number $q$ is 6. We will specifically study an antiferromagnetic
Ising model on them and solve it exactly to investigate various equilibrium and
time-independent metastable phases. The methods we introduced and studied in
this paper may provide an alternative approach, which is associated with easier
calculation and much less time-consuming than typical simulation methods such as
MC or MD, to investigate the ideal glass transition of small molecular systems in 3-D.

\section{Recursive Lattice Geometry}

\subsection{Recursive Lattice Approximation}

Except in some particular cases \cite{Ising,Onsager,Ohzeki}, a many-body system with interactions
on a regular lattice is difficult to be solved exactly because of the
complexity involved with treating the combinatorics generated by the
interaction terms in the Hamiltonian when summing over all states. Usually the
mean-field approximation is adopted to handle this problem. On the other hand,
recursive lattices enable us the explicit treatment of combinatorics on these
lattices and no approximation is necessary. The recursive lattice
is chosen to have the same coordination number $q$ of the regular lattice, which it is
designed to describe. As usual, $q$ is the number of edges connected to a vertex. 

\begin{figure}
[ptb]
\begin{center}
\includegraphics[width=0.5\textwidth]{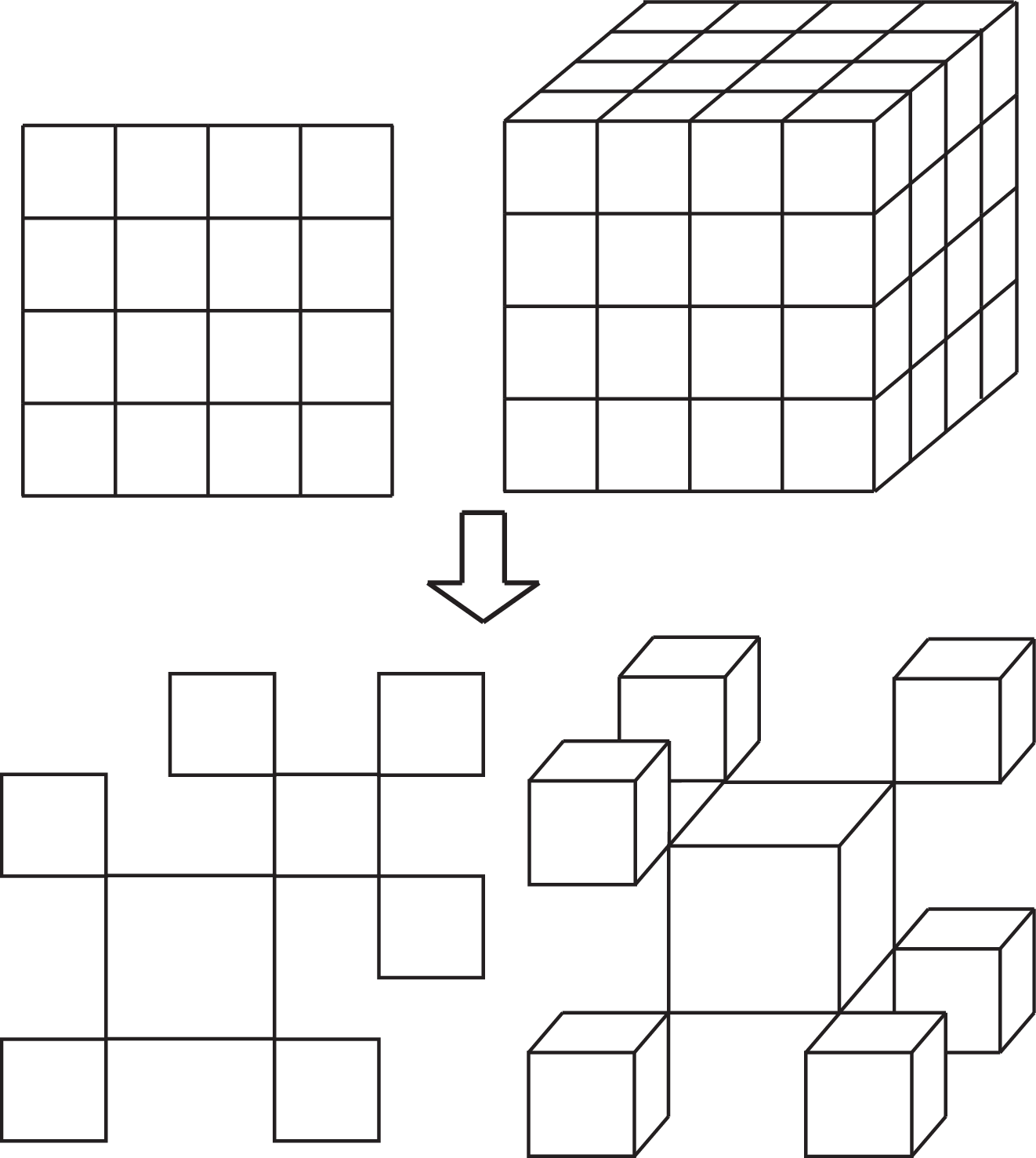}
\caption{Husimi lattice (HL) and the cubic recursive lattice (CRL) constructed from the basic units of regular lattices.}
\label{fig1}
\end{center}
\end{figure}

The original HL and the CRL as the analogs of the 2-D and 3-D lattices are shown in Fig. \ref{fig1}. The coordination number in HL is 4, which is the same as it is in a regular square lattice (2-D case), while in CRL $q$ is 6 to match the regular cubic lattice (3-D case). With the identical $q$, The recursive lattice calculations have been demonstrated to be highly reliable approximations to the regular lattices \cite{pdg_prl_reliable}. The most impressive property of the recursive lattices is that, a certain number of independent lattice units are joint at one vertex to give rise to a tree-like
structure. This independence of units enables various manipulations on the structures, which is impossible in the regular lattices. We can use any kind of artificial geometrical structure,
called a unit cell, containing any number of sites to construct a recursive
lattice with the provision that one unit is only connected to the neighbor unit
by a shared vertex, and the neighbor unit connects to the next unit by
the next shared site, and so on in a recursive fashion. 

\subsection{Multi-branched Recursive Lattice}

\begin{figure}
[ptb]
\begin{center}
\includegraphics[width=0.5\textwidth]{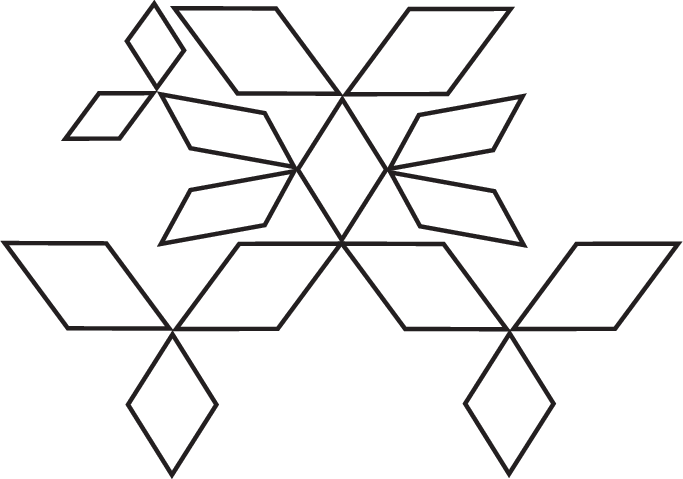}
\caption{A 3-branched Husimi lattice with $q=6$. Note that
all the units are square and equal in size, however they cannot be drawn in that way
on page.}%
\label{fig2}%
\end{center}
\end{figure}

As mentioned above, the HL with $q=4$ is formed by
connecting two square units of the regular square lattice of the same $q$, and it is the same for the CRL of $q=6$ to describe the regular cubic lattice. This
may cause a misunderstanding that the recursive lattice of a given
coordination number $q$ can only be constructed from the single unit of a
regular lattice with that particular $q$, which is not true. It should be recognized that the
coordination number alone does not determine the structure of basic unit. For
example, we can modify the original HL by joining three square units
at one vertex, as shown in Fig. \ref{fig2}, to result in $q=6$. This is made possible by the independence of units in recursive lattice. In this case, we may ask the following
questions: Is this modified HL, with three branches joint on one vertex, also capable to approximate a 3-D
regular lattice like the CRL based on the same $q$? If yes, what
advantages or disadvantages does it hold, compared to the CRL? These questions
will be investigated in this work.%

\section{Antiferromagnetic Ising Model}

An Ising spin $S$ has two possible values, which usually are assigned as $+1$
(up or $+$) and $-1$ (down or $-$) \cite{Ising, Onsager, Wu}, and they can represent a
particle and the void in a monatomic system, or particle A and B in an alloy. The Ising model with the nearest-neighbor interactions between neighboring spins $S_{i}$ and
$S_{j}$ and in the presence of an external magnetic field $H$ has the energy
given by
\[
E=-J{\sum_{<i,j>}}S_{i}S_{j}-H{\sum}S_{i},
\]
where $<i,j>$ is the nearest-neighbor pairs of lattice sites $i$ and $j$, and
$S_{i}=\pm1$ is the spin on the site $i$. The value of $J$ can be setup to be either positive or negative. For the $J>0$ case, the spins in the same state (the parallel arrangement) will have the lowest energy at absolute zero, so that the model has a ferromagnetic ordering. Vice versa, the anti-parallel arrangement at absolute zero is favored by the antiferromagnetic interaction
$J<0$. 

We will take the antiferromagnetic interaction in this paper for two reasons: (1) The ferromagnetic case prefers only one spin state aggregated together at the near-zero temperature, which implies a total phase separation of two particles A and B in the alloy case, or a compressible pure system ruling out all the voids. On the other hand, the antiferromagnetic case provide a uniform state at low temperature, which is facilitated for us to investigate the equilibrium thermodynamics; (2) While the ferromagnetic case has only one solution, the antiferromagnetic interaction can provide two solutions representing the stable and metastable states, which is necessary to study the supercooled liquid and glass transition. The details of this so-called ``solutions" will be discussed later.

The standard Ising model only includes the interaction between the nearest
neighbor spins. Nevertheless, it is not realistic that particles farther away than
the nearest distance do not interact with each other in real systems. For a
better approximation of a real system, more interaction energy terms such as
the interactions between second neighbor sites, interaction between three and
four spins are also considered in our model.

\section{Model, Fix-point solutions and Thermodynamics}

The calculation of HL and CRL are similar in principle. As a basic
unit cell at a higher level shares a single site connecting it to the lower
level, this site in each unit plays a role different from the other sites. We
will call this the base site and the corresponding spin at that site will be
generically denoted by $S^{\text{b}}$. In particular, we can define the energy
$e_{\alpha}$ associated with a cell $\alpha$ by paying special attention to
this site in such a way that the sum over all cells will give the energy $E$
of the model over the entire recursive lattice. Here we will take the
multi-branched HL for example to demonstrate the calculation
method, and briefly give the generalized formula for CRL, for which the
details of calculations were described in our previous works, e. g. the Ref. \cite{PDG4}. We will, however, present all relevant details here.

\subsection{Spins Conformation}

\begin{figure}
[ptb]
\begin{center}
\subfloat[]
{\includegraphics[width=0.5\textwidth]{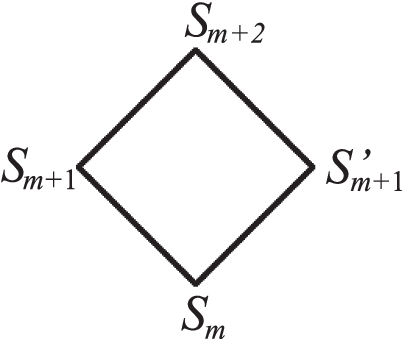}
}\\
\subfloat[]
{\includegraphics[width=0.5\textwidth]{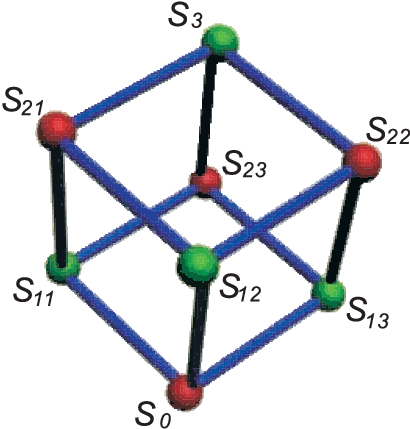}
}
\caption{Level index and sites labeling in the (a) square and (b) cubic unit.}%
\label{fig3}
\end{center}
\end{figure}

We first consider a HL made up by squares as the unit cells. The
four spins in such a square with its base site at the $m$-th level are shown
in Fig. \ref{fig3}a, with $S_{m}$ as the base site. The peak site is at level
$m+2$ and the remaining intermediate sites are at level $m+1$. The spins at
the intermediate sites are denoted by $S_{m+1}$ and $S_{m+1}^{\prime}$,
respectively. The energetic of the model is defined by considering these
squares. We now introduce following quantities for a square cell:
\begin{align*}
A_{\text{mag}}  &  =S_{m+1}+S_{m+1}^{\prime}+S_{m+2};\\
A_{\text{nb}}  &  =S_{m}S_{m+1}+S_{m}S_{m+1}^{\prime}+S_{m+1}S_{m+2}%
+S_{m+1}^{\prime}S_{m+2};\\
A_{\text{sd}}  &  =S_{m}S_{m+2}+S_{m+1}S_{m+1}^{\prime};\\
A_{\text{tri}}  &  =S_{m}S_{m+1}S_{m+1}^{\prime}+S_{m}S_{m+1}^{\prime}%
S_{m+2}+S_{m}S_{m+1}S_{m+2}+S_{m+1}S_{m+1}^{\prime}S_{m+2};\\
A_{\text{qua}}  &  =S_{m}S_{m+1}S_{m+1}^{\prime}S_{m+2}.
\end{align*}

Note that the base site is not included in the magnetic term $A_{\text{mag}}$;
instead we only count the contributions from the sites of the unit cell above
the site $S_{m}$. The missing contribution from this site will be accounted in the next unit on lower level, when we construct the recursion relations below.

Let $\gamma$ denote a particular spin state in the cell. The energy of the
Ising model in the cell for a particular cell state $\gamma$ is:%

\begin{equation}
e(\gamma)=-JA_{\text{nb}}(\gamma)-J_{P}A_{\text{sd}}(\gamma)-J^{\prime
}A_{\text{tri}}(\gamma)-J^{\prime\prime}A_{\text{qua}}(\gamma)-HA_{\text{mag}%
}(\gamma), \label{e_alpha}%
\end{equation}
where $J_{P}$ is the interaction energy between the second nearest sites,
$J^{\prime}$ is the interaction energy of three sites (triplet), and
$J^{\prime\prime}$ is the interaction energy of the four sites (quadruplet).

The total energy of the Ising model mentioned above is the sum of the energy
of all cells $\alpha$\ on our recursive lattice:%

\begin{equation}
E(\Gamma)=%
{\textstyle\sum_{\alpha}}
e(\gamma_{\alpha})-HS_{0}, \label{Model_Energy}%
\end{equation}
where $\Gamma=%
{\textstyle\bigotimes\nolimits_{\alpha}}
\gamma_{\alpha}$ denotes the spin state of the lattice, and $S_{0}$ now
represents the spin at the origin of the recursive lattice. It should be
addressed here that, since our lattices are homogeneous and infinitely large,
any site on the lattice can be artificially selected to be the origin site.
There are four sites in a square cell; hence, the number of possible states of
a cell is $2^{4}=16$. The 1st to the 8th states correspond to the base spin
$S_{m}=+1$, and the 9th to the 16th states to $S_{m}=-1$ on the base site. The
number of lattice conformations $\Gamma$ for a lattice with $N_{\text{C}}$
cells is $16^{N_{\text{C}}/r}$, where the coefficient $r$ is the number of
branches joined at one site for a multi-branched lattice.

Note that in this paper the temperature $T$ and $\beta=1/T$ are normalized with Boltzmann constant
$k_{B}=1$. The Boltzmann weights are given by

\begin{equation}
w(\gamma)=exp(-\beta e(\gamma)), \  
w(\Gamma)=\exp(\beta HS_{0}){\textstyle\prod\nolimits_{\alpha}}w(\gamma_{\alpha}). \label{weights}%
\end{equation}
Then the partition function (PF) is a sum over lattice configurations:%

\begin{equation}
Z(T)=\underset{\Gamma}{\sum}w(\Gamma). \label{PF}%
\end{equation}

By choosing proper values of the interaction energies, we can simulate various
systems to study their thermodynamics and phase transitions under
different conditions. As a convention, we take $J=-1$ to set the temperature
scale for our model.

\subsection{Fix-point solution}%

\begin{figure}
[ptb]
\begin{center}
\includegraphics[width=0.5\textwidth]{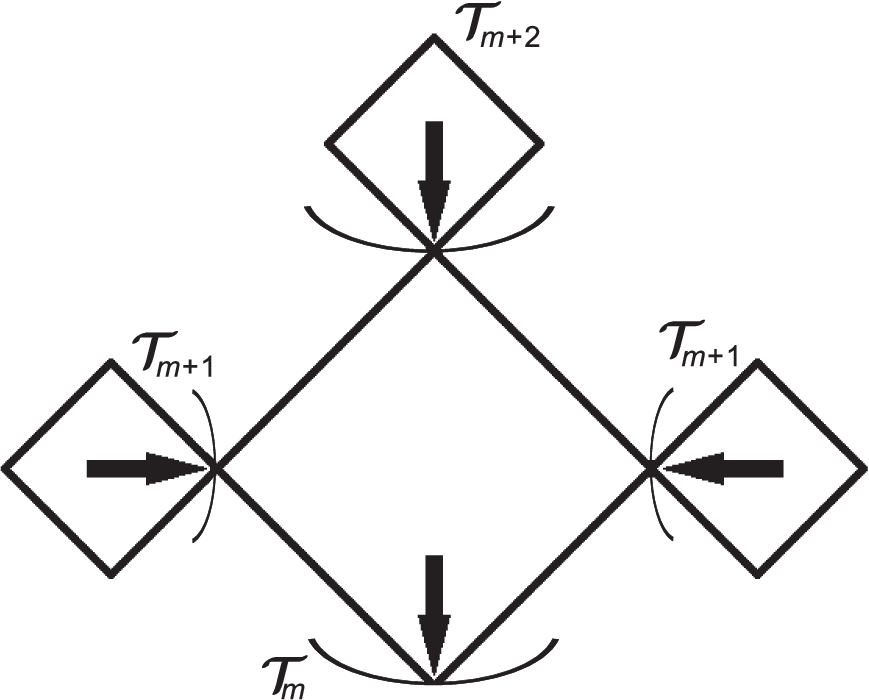}%
\caption{Contributions of sub-branches in the HL.}%
\label{fig4}%
\end{center}
\end{figure}

Since the units are connected together only at a site, the units on higher
levels from some site may be treated as a branch linked to a particular site. The branch
$\mathcal{T}_{m}$ is obtained by cutting the lattice at the $m$-th site; see in 
Fig. \ref{fig4}. This branch is said to have its base site at the $m$-th
level. A partial partition function (PPF) $Z_{m+1}(S_{m+1})$ for the branch
$\mathcal{T}_{m+1}$ with its base at the level $m+1$ can be introduced to
represent the contribution of the branch with spin $S_{m+1}$ fixed as its base site.
Then the PPF $Z_{m}(S_{m})$ at the lower level $m$ can be expressed in terms
of PPF's $Z_{m+1}(S_{m+1})$, $Z_{m+1}(S_{m+1}^{\prime})$, and $Z_{m+2}(S_{m+2})$
at higher levels. Thus, recursively we can start from an initial site to calculate the contribution of the entire lattice towards the origin. In this way, we can obtain the thermodynamics of the entire system.

The partition function $Z_{0}$ of the lattice can be represented by the
contributions to the origin site from the $r$ branches $T_{0}$ meeting at the
origin with either $S_{0}$:%

\begin{equation}
Z_{0}=Z_{0}^{r}(+)e^{\beta H}+Z_{0}^{r}(-)e^{-\beta H} \label{PF_Origin}%
\end{equation}
The weights of the magnetic field above is due to the contribution from
the base site, which is not included in $A_{\text{mag}}$.

We now turn to the PPF's. As each PPF at a base site in the square is a sum of
the $8$ configurations of the remaining three sites in the square, with
$r^{\prime}\equiv r-1$ we have

\begin{align}
Z_{m}(+)  &  =\underset{\gamma=1}{\overset{8}{\sum}}Z_{m+1}^{r^{\prime}%
}(S_{m+1})Z_{m+1}^{r^{\prime}}(S_{m+1}^{\prime})Z_{m+2}^{r^{\prime}}%
(S_{m+2})w(\gamma),\label{PPF+_Recursion}\\
Z_{m}(-)  &  =\underset{\gamma=9}{\overset{16}{\sum}}Z_{m+1}^{r^{\prime}%
}(S_{m+1})Z_{m+1}^{r^{\prime}}(S_{m+1}^{\prime})Z_{m+2}^{r^{\prime}}%
(S_{m+2})w(\gamma), \label{PPF-_Recursion}%
\end{align}
with a product from the $r^{\prime}$ branches at the two sites on level $m+1$
and the $r^{\prime}$ branches at the peak site at $m+2$.

We now introduce the ratios

\begin{equation}
x_{m}=\frac{Z_{m}(+)}{Z_{m}(+)+Z_{m}(-)}, \ y_{m}=\frac{Z_{m}(-)}{Z_{m}(+)+Z_{m}(-)}. \label{Ratios}%
\end{equation}
and

\begin{equation}
z_{m}(S_{m})=\left\{
\begin{array}
[c]{c}%
x_{m}\text{ if }S_{m}=+1\\
y_{m}\text{ if }S_{m}=-1
\end{array}
\right.  \label{General Ratio}%
\end{equation}
In terms of
\[
B_{m}^{r\prime}=Z_{m}(+)+Z_{m}(-),
\]
we have $Z_{m}(+)=B_{m}^{r\prime}x_{m}$ and $Z_{m}(-)=B_{m}^{r\prime}y_{m}$.
By use of (\ref{General Ratio}), we obtain
\begin{align*}
B_{m}^{r\prime}z_{m}(\pm)  &  =\sum B_{m+1}^{r\prime}z_{m+1}^{r^{\prime}%
}(S_{m+1})B_{m+1}^{r\prime}z_{m+1}^{r^{\prime}}(S_{m+1}^{\prime}%
)B_{m+2}^{r\prime}z_{m+2}^{r^{\prime}}(S_{m+2})w(\gamma),\\
z_{m}(\pm)  &  =\sum z_{m+1}^{r^{\prime}}(S_{m+1})z_{m+1}^{r^{\prime}}%
(S_{m+1}^{\prime})z_{m+2}^{r^{\prime}}(S_{m+2})w(\gamma)/Q(x_{m+1},x_{m+2}),
\end{align*}
where the sum is over $\gamma=1$, $ 2 $, $ 3 $, \ldots, $8$ for $S_{m}=+1$, and over
$\gamma=9$, $10$, $11$, \ldots, $16$ for $S_{m}=-1$, and where
\[
Q(x_{m+1},x_{m+2})\equiv\left[ B_{m}/B_{m+1}^{2}B_{m+2}\right]  ^{r^{\prime}};
\]
it is related to the polynomials
\begin{align*}
Q_{+}(x_{m+1},x_{m+2})  &  =\underset{\gamma=1}{\overset{8}{\sum}}%
z_{m+1}^{r^{\prime}}(S_{m+1})z_{m+1}^{r^{\prime}}(S_{m+1}^{\prime}%
)z_{m+2}^{r^{\prime}}(S_{m+2})w(\gamma),\\
Q_{-}(x_{m+1},x_{m+2})  &  =\underset{\gamma=9}{\overset{16}{\sum}}%
z_{m+1}^{r^{\prime}}(S_{m+1})z_{m+1}^{r^{\prime}}(S_{m+1}^{\prime}%
)z_{m+2}^{r^{\prime}}(S_{m+2})w(\gamma),
\end{align*}
according to

\[
Q(x_{m+1},x_{m+2})=Q_{+}(x_{m+1},x_{m+2})+Q_{-}(y_{m+1},y_{m+2}).
\]
In terms of the above polynomials, we can express the recursive relation for
the ratio $x_{m}$ in terms of $x_{m+1}$ and $x_{m+2}$:%

\begin{equation}
x_{m}=\frac{Q_{+}(x_{m+1},x_{m+2})}{Q(x_{m+1},x_{m+2})}.
\label{Ratios_Polynomial_x}%
\end{equation}
A similar relation holds for $y_{m}\equiv1-x_{m}$:%

\begin{equation}
y_{m}=\frac{Q_{-}(y_{m+1},y_{m+2})}{Q(y_{m+1},y_{m+2})}.
\label{Ratios_Polynomial_y}%
\end{equation}

Clearly from the definition of $x$, the physical meaning of the solution is the ratio of sub-trees' thermodynamic contributions to a particular site (excluded), i.e. the cavity magnetization \cite{cavity}, therefore it determines the probability that a site is occupied by the spin $S=+1$, and we can subsequently take it to be the solution of the entire lattice. Because of the recursive property of these equations, we would expect the cycle-form solutions, which is called the \emph{fix-point} (FP) solution. Then the bulk thermodynamic calculation just focuses on solving the Eqs. (\ref{Ratios_Polynomial_x}, \ref{Ratios_Polynomial_y}) and determine the FP solution. In this work, our calculation provides two types of solutions for each model, the alternating state solution (2-cycle) and the uniform solution (1-cycle). At high temperatures, we find a uniform FP solution $x$ on all the sites, which we call the 1-cycle solution:

\[
x=\frac{Q_{+}(x,x)}{Q(x,x)}.
\]

At low temperatures, we find alternating solutions $x_{1}$ and $x_{2}$ on the two
successive levels, which we call a 2-cycle solution:%

\[
x_{1}=\frac{Q_{+}(x_{2},x_{1})}{Q(x_{2},x_{1})}, \ x_{2}=\frac{Q_{+}(x_{1},x_{2})}{Q(x_{1},x_{2})}.
\]

This describes the antiferromagnetic ordering, i.e. two neighboring spins prefer to be in different states at low temperature. In contrast, the 1-cycle solution represents the less-ordered state, in which all the spins have the same probability to be in either states. Based on the nature of antiferromagnetic Ising model, a three or higher cycle solutions is hard to imagine. And although
we cannot prove that there is no solution other than the 1- and 2-cycle form, we have not found any. The details of solutions will be discussed in the later section. 

\subsection{Calculation of thermodynamics}

\subsubsection{Free Energy}

It is clear that the 1-cycle solution is a special case of the 2-cycle
solution with $x_{1}=x_{2}$. Therefore, we will assume a 2-cycle solution and
determine its thermodynamics in this section. If it happens that $x_{1}=x_{2}$, we obtain the
1-cycle thermals. The solution with a higher free energy will then describe
a metastable state, which will be useful to identify the supercooled
liquid and the glass transition.

\begin{figure}
[ptb]
\begin{center}
\includegraphics[width=0.5\textwidth]{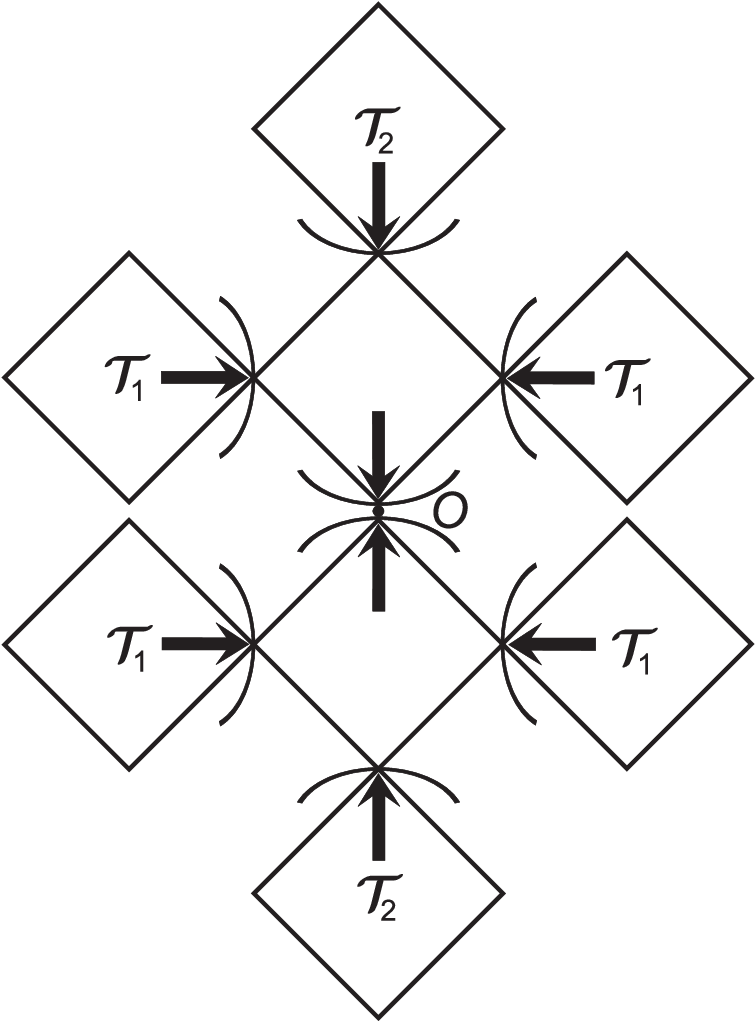}%
\caption{Two squares meeting at the origin of the lattice. In general, there
are $r$ units meeting at the origin.}%
\label{fig5}%
\end{center}
\end{figure}

The Helmholtz free energy is given by

\[
F=-T\log Z
\]
and requires obtaining the partition function $Z$, which is given by Eq. (\ref{PF_Origin}). As the lattice is
infinitely large, there is no sense in calculating $Z$ or $F$, each of which
will also be infinite. At the FP solution, we can use its homogeneity over
two levels to obtain the free energy per site by following the Gujrati trick
[11], which we briefly describe below. We consider the HL made up
by connecting $r$ squares at each site. The partition function of the whole
system is given by considering the $r$ branches $\mathcal{T}_{0}$ meeting at
the origin as shown in Fig. \ref{fig5}. 

Imagine that we cut off the $rr^{\prime}$ branches $\mathcal{T}_{2}$ on
level 2 and hook up them together to form $r^{\prime}$ smaller lattices. The
partition function of each of these smaller lattices is:%

\[
Z_{2}=Z_{2}^{r}(+)e^{\beta H}+Z_{2}^{r}(-)e^{-\beta H}.
\]
Similarly we can imagine to cut off the $2rr^{\prime}$ branches $\mathcal{T}%
_{1}$ and hook them up to form $2r^{\prime}$ smaller lattices. The partition
function of each smaller lattice is%

\[
Z_{1}=Z_{1}^{r}(+)e^{\beta H}+Z_{1}^{r}(-)e^{-\beta H}.
\]

Since free energy is an extensive quantity, the free energy of the whole
system is the sum of the free energies of the smaller lattices and the $r$
local squares that are left out after dissecting the original lattice. The
free energy of the left out squares is

\[
F_{local}=-T\log\left[  \frac{Z_{0}}{(Z_{1}^{2}Z_{2})^{r^{\prime}}}\right]  .
\]
There are $4/r$ cites per square and $r$ squares in the local region. Thus,
the free energy per site is:%

\begin{equation}
F=-\frac{F_{local}}{4}. \label{FreeEnergy/site}%
\end{equation}
By substituting $Z_{m}(+)=B_{m}x_{m}$ and $Z_{m}(-)=B_{m}y_{m},$ we finally obtain%

\begin{equation}
F=-\frac{1}{4}T\log(Q^{2r^{\prime}}\frac{1}{\{[x_{1}^{r}e^{\beta H}%
+(1-x_{1})^{r}e^{-\beta H}]^{2}[x_{2}^{r}e^{\beta H}+(1-x_{2})^{r}e^{-\beta
H}]\}^{r^{\prime}}}) \label{FreeEnergyPolynomial/site}%
\end{equation}
Using the FP solution $x_{1}$ and $x_{2}$ from the FP calculation, the
numerical value of free energy can be easily achieved.%

\subsubsection{Conformation Probability}

A square unit with 4 spins on each site has sixteen possible conformations
$\gamma$. The probability of a given conformation $\gamma$ is
\[
P(\gamma)=\frac{\left[  Z_{0}(S_{0})Z_{1}(S_{1})Z_{1}(S_{1}^{\prime}%
)Z_{2}(S_{2})\right]  ^{r^{\prime}}\exp(\beta HS_{0})w(\gamma)}{Z_{0}},
\]
which satisfies the sum rule%
\[
\overset{16}{\underset{\Gamma}{\sum}}P(\gamma)=1.
\]
It is easy to obtain that%
\[
P(\gamma)=\frac{\left[  z_{0}(S_{0})z_{1}(S_{1})z_{1}(S_{1})z_{2}%
(S_{2})\right]  ^{r^{\prime}}\exp(\beta HS_{0})w(\gamma)}{Q_{_{+}}x_{0}%
^{r}e^{\beta H}+Q_{_{-}}(1-x_{0})^{r}e^{-\beta H}}%
\]

\subsubsection{Energy Density}

The energy density is defined as the summation over the product of the energy
and the probability of a conformation state. Once we have the conformation
probabilities $P(\gamma)$, energy density (per site) can be simply calculated by
evaluating the sum of the product of energy of each conformation and the
$P(\gamma)$:%
\[
E=\frac{r}{4}%
{\textstyle\sum}
e(\gamma)P(\gamma)
\]
where%
\[
e_{\alpha}=-JA_{nb}-J_{P}A_{\text{sd}}-J^{\prime}A_{\text{tri}}-J^{\prime
\prime}A_{\text{qua}}-H(S_{\alpha}^{\text{b}}+A_{\text{mag}})/r
\]
is the energy of a unit cell $\alpha$ in a configuration $\gamma$ as mentioned
before. The base site $S_{\alpha}^{\text{b}}$ is included in the magnetic term
as it contributes to the energy of each cell.

\subsubsection{Entropy}

With the free energy and energy density, by definition we
simply have the entropy as%
\[
S=\beta(E-F),
\]
or from the first derivative of the free energy with respect to the temperature:%
\[
S=-\partial F/\partial T.
\]
Depends on the setup of energy parameters, the entropy obtained from 1-cycle
solution is possible to be zero at some temperature $T_{\text{K}}$ below the
melting transition temperature $T_{\text{M}}$, since it represents the
metastable supercooled liquid. The entropy will become negative below
$T_{\text{K}}$ with continuous cooling. However negative entropy is known as
unphysical in nature, thus the metastable state must undergo a transition and
change to the glass state before $T_{\text{K}}$, which in this way is defined as
the ideal glass transition temperature.

\subsection{The CRL calculation}

\begin{figure}
[ptb]
\begin{center}
\includegraphics[width=0.5\textwidth]{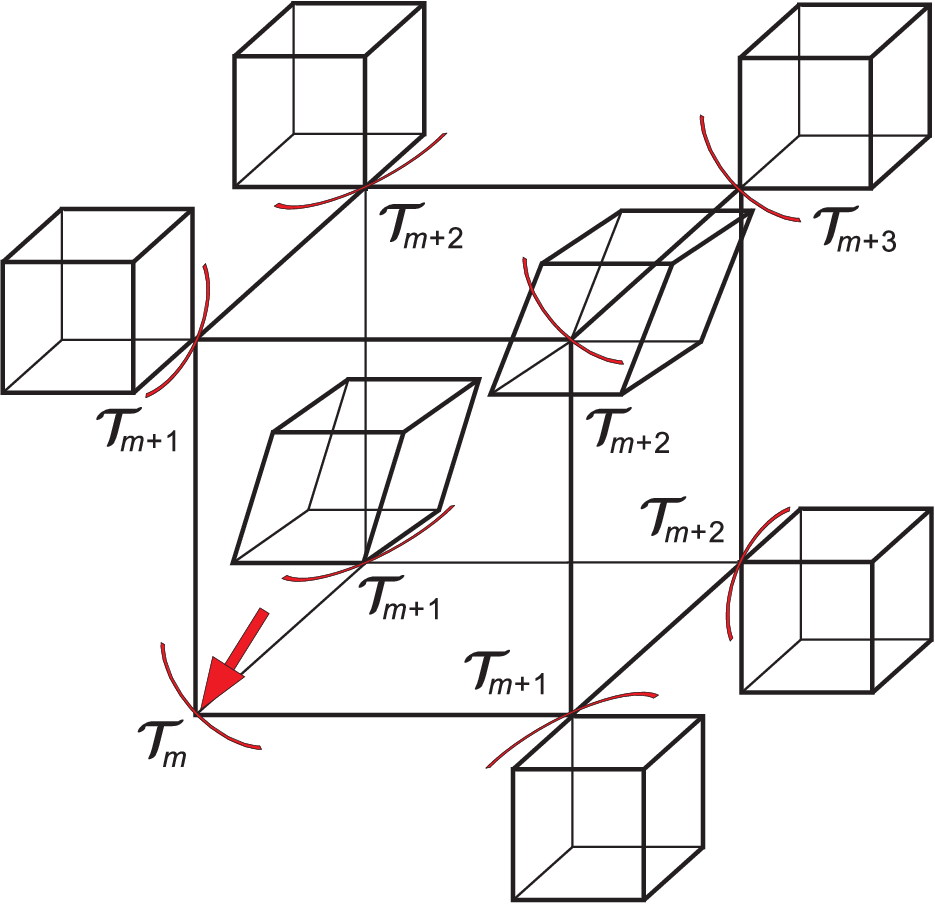}%
\caption{The sub-tree contributions in the CRL.}%
\label{fig6}
\end{center}
\end{figure}

The CRL calculation is similar to the method introduced in the previous section
and can be found in Ref. \cite{Ran1}. Here we briefly give the scheme of the CRL
calculation. In a CRL the sites on a cube unit are divided
into four layers and labeled as follows: $S_{m}$, $(S_{m+1,1}$, $S_{m+1,2}$
, $S_{m+1,3})$, $(S_{m+2,1}$, $S_{m+2,2}$, $S_{m+2,3})$, and $S_{m+3}$; see Fig. \ref{fig3}b
where we take $m=0$. The sub-tree contributions in the CRL are demonstrated in Fig. \ref{fig6}.

Taking their contributions at each site easily gives the recursion relations
for the PPFs:

\begin{align*}
Z_{m}(+)  &  =\underset{\gamma=1}{\overset{128}{\sum}}[\overset{3}%
{\underset{\{m+1\}}{\prod}}Z_{m+1}(S_{m+1})][\overset{3}{\underset
{\{m+2\}}{\prod}}Z_{m+2}(S_{m+2})]Z_{m+3}(S_{m+3})w(\gamma),\\
Z_{m}(-)  &  =\underset{\gamma=129}{\overset{256}{\sum}}[\overset{3}%
{\underset{\{m+1\}}{\prod}}Z_{m+1}(S_{m+1})][\overset{3}{\underset
{\{m+2\}}{\prod}}Z_{m+2}(S_{m+2})]Z_{m+3}(S_{m+3})w(\gamma),
\end{align*}

By introducing $B_{m}$ and the ratios $x_{m}$, $y_{m}$, and $z_{m}%
(S_{m})$ (Eq. \ref{General Ratio}) as before, we have%

\begin{align*}
Q_{+}(x_{m+1},x_{m+2},x_{m+3})  &  =\underset{\gamma=1}{\overset{128}{\sum}%
}[\overset{3}{\underset{\{m+1\}}{\prod}}z_{m+1}(S_{m+1})][\overset
{3}{\underset{\{m+2\}}{\prod}}z_{m+2}(S_{m+2})]z_{m+3}(S_{m+3})w(\gamma),\\
Q_{-}(x_{m+1},x_{m+2},x_{m+3})  &  =\underset{\gamma=129}{\overset{256}{\sum}%
}[\overset{3}{\underset{\{m+1\}}{\prod}}z_{m+1}(S_{m+1})][\overset
{3}{\underset{\{m+2\}}{\prod}}z_{m+2}(S_{m+2})]z_{m+3}(S_{m+3})w(\gamma),
\end{align*}
and
\[
Q(x_{m+1},x_{m+2},x_{m+3})=Q_{+}(x_{m+1},x_{m+2},x_{m+3})+Q_{-}(y_{m+1}%
,y_{m+2},y_{m+3})=\frac{B_{m}}{B_{m+1}^{3}B_{m+2}^{3}B_{m+3}}.
\]
We can now obtain the recursive relation for the ratio $x_{m}$ in terms of
$x_{m+1}$, $x_{m+2}$ and $x_{m+3}$:%
\[
x_{m}=\frac{Q_{+}(x_{m+1},x_{m+2},x_{m+3})}{Q(x_{m+1},x_{m+2},x_{m+3})}.
\]
A similar relation holds for $y_{m}\equiv1-x_{m}$:%
\[
y_{m}=\frac{Q_{-}(y_{m+1},y_{m+2},y_{m+3})}{Q(y_{m+1},y_{m+2},y_{m+3})}.
\]

For the CRL case, we also have the 1-cycle solution, $x_{m}=x_{m+1}%
=x_{m+2}=x_{m+3}$, which represents the disordered state, and the 2-cycle
solution, $x_{m}=x_{m+2}$ and $x_{m+1}=x_{m+3}$, which represents the ordered
stable state. Thermodynamic properties such as free energy and entropy can be
calculated from the solutions using similar techniques as before. The local
area around the origin is chosen to be the two cubes joint on the origin
site, and it has eight sites. The final expression of the free energy per
site is:
\begin{equation}
F=\frac{1}{8}F_{local}=-\frac{1}{8}T\log(Q^{2}\frac{1}{[x_{1}^{2}e^{\beta
H}+(1-x_{1})^{2}e^{-\beta H}]^{4}[x_{2}^{2}e^{\beta H}+(1-x_{2})^{2}e^{-\beta
H}]^{2}}) \label{CRLFreeEnergy/site}%
\end{equation}

\section{Results and Discussion}

The thermodynamic calculations have been done on 3-branched HL ($r=3$) and
CRL. Since both models are to approximate the 3-D regular lattice, the results
will be compared in this section. The role of energy parameters will be
studied in both cases. In the following discussion, we will call the parameters setup of $J=-1$ and all other parameter to be zero as the \emph{reference case}.

\subsection{General solutions with $J=-1$ and all other parameter are zero}
\begin{figure}
[ptb]
\begin{center}
\includegraphics[width=0.8\textwidth]{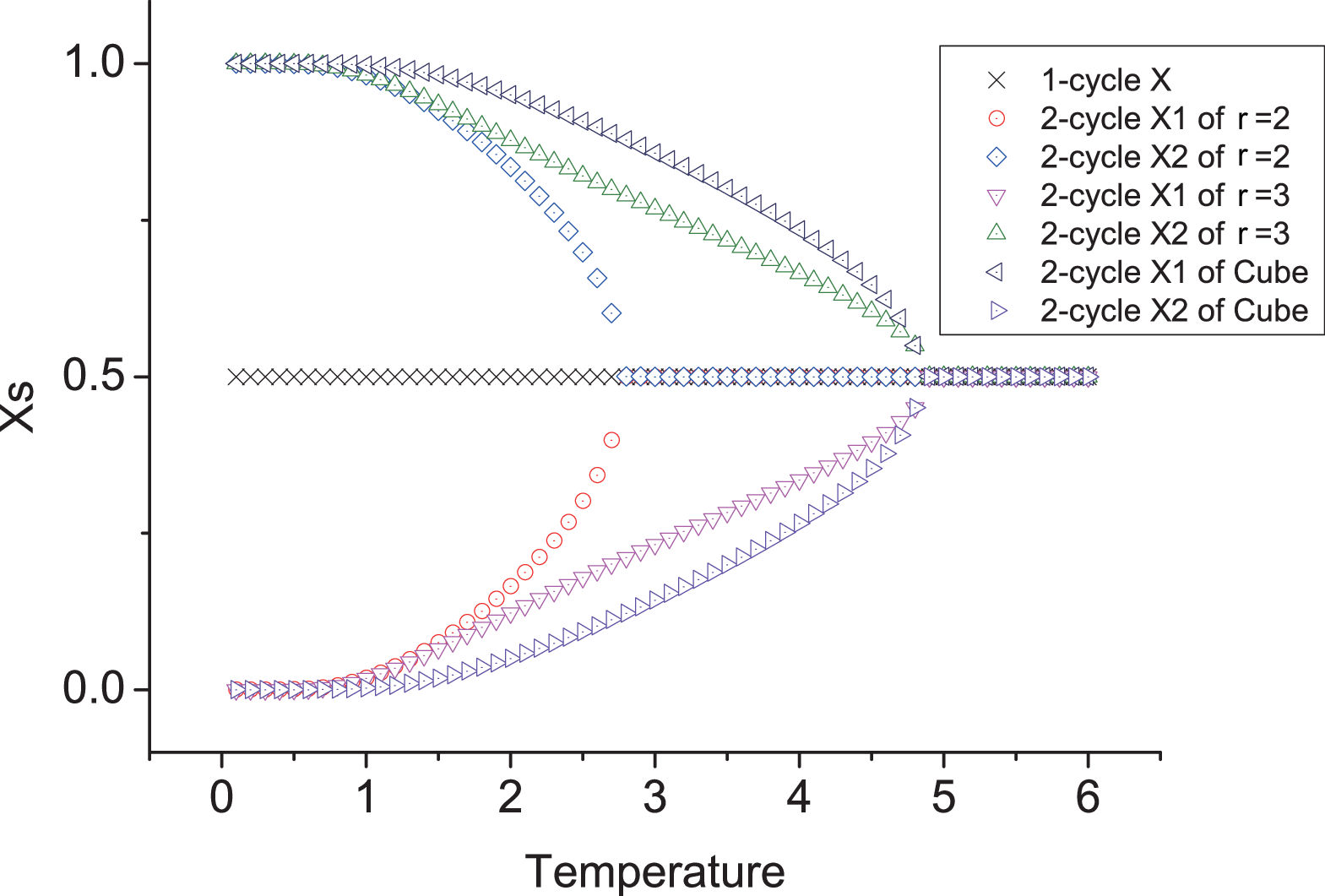}
\caption{The reference 2-cycle and the 1-cycle solutions of original HL, HL of $r=3$, and CRL ($J=-1$ and other parameters are zero).}
\label{fig7}%
\end{center}
\end{figure}

The reference 1- and 2-cycle solutions of the original HL, HL of $r=3$, and CRL are shown in Fig. \ref{fig7}. We can see that the 1-cycle solutions of all the models are $x=0.5$ everywhere, which implies that the probability of a site occupied by $S=+1$ is simply fifty percent with the absence of magnetic field $H$ in the Hamiltonian. This situation corresponds to a
disordered phase and represents a high free energy state at low temperature
(the supercooled liquid) \cite{23}. For the 2-cycle solutions, one branch
approaches to $1$ and the other approaches to $0$ as $T\rightarrow0$,
thus we have $S=+1$ and $S=-1$ alternatively occupying the two neighboring sites at low $T$. This structure represents the ordered states (crystal) and is characterized by the complete antiferromagnetic order in the Ising model.

To clarify more about the physical meaning the solutions in our model: The 2-cycle solution is the most ordered state at a particular temperature, subsequently it is the crystal; Meanwhile, the 1-cycle is the ``most disordered state", i.e. an antiferromagnetic Ising model cannot be more disordered with fifty percent probability of spins states everywhere. However, since it is still an achievable solution, it should be taken as the upper limit of the amorphous states, i.e. the most unstable state out of all the possible states. In this way, any artificial values between 0 and 1 of solutions $x_{1}$ and $x_{2}$ with summation of 1 correspond to a possible state of the system. In the next section, the thermodynamics of 1-cycle solution will be discussed and it behaves exactly as the supercooled liquid and exhibits the Kauzmann paradox, which agrees with our expectation that this less stable solution is the metastable one.

At a critical temperature $T=2.75$, the 2-cycle solution turns to be 1-cycle and
there occurs a transition for the original HL,
where the crystal state turns into the disordered state. This represents the melting transition $T_{\text{M}}$ \cite{24,25} from crystal to liquid. For the HL of $r=3$, the melting temperature increases significantly and becomes
very close to the one in CRL. The $T_{\text{M}}$ is found to be 4.84 in
CRL and 4.89 in the 3-branched HL. These values are comparable to the
results obtained by other methods on the Ising antiferromagnetic model on a
simple cubic lattice: 4.51 by Monte Carlo \cite{26}, 4.93 by mean field
renormalization group \cite{27} and 4.52 by series expansion \cite{28}. This adds faith that
our model is not far from a realistic model. We can observe that the singular temperature moves towards the "correct" temperature 4.51 or
4.52 obtained by simulation or series analysis as we increase the coordination
number in the case of the two HLs, or use a more realistic unit
cell (the cube). Meanwhile, they drift away from the mean field renormalization group
result.

The difference between the 2-cycle solutions in both lattices is also
remarkable. It is as expected that HL of $r=3$ is better to
describe the 3-D case than the HL of $r=2$. However, in Fig. \ref{fig7}, the curves of $x$s of the $r=3$ HL are more like the stretched curves of the original HL solutions, while the $x$s of CRL are more rounded curves. The difference between $x_{0}$ and $x_{1}$ is larger in CRL. Since the 1 and 0 cyclic solution at the zero temperature corresponds to the ideal
crystal stable state, a 2-cycle can be identified to be more stable if its two
branches are closer to 1 and 0 respectively, by this meaning the 2-cycle solution
in CRL is more stable below $T_{\text{M}}$ than in the
3-branched HL. This difference can also be observed in the thermal
properties discussed below.

\subsection{Thermodynamics of $r=3$ HL and CRL with reference case}

\begin{figure}
[ptb]
\begin{center}
\includegraphics[width=0.8\textwidth]{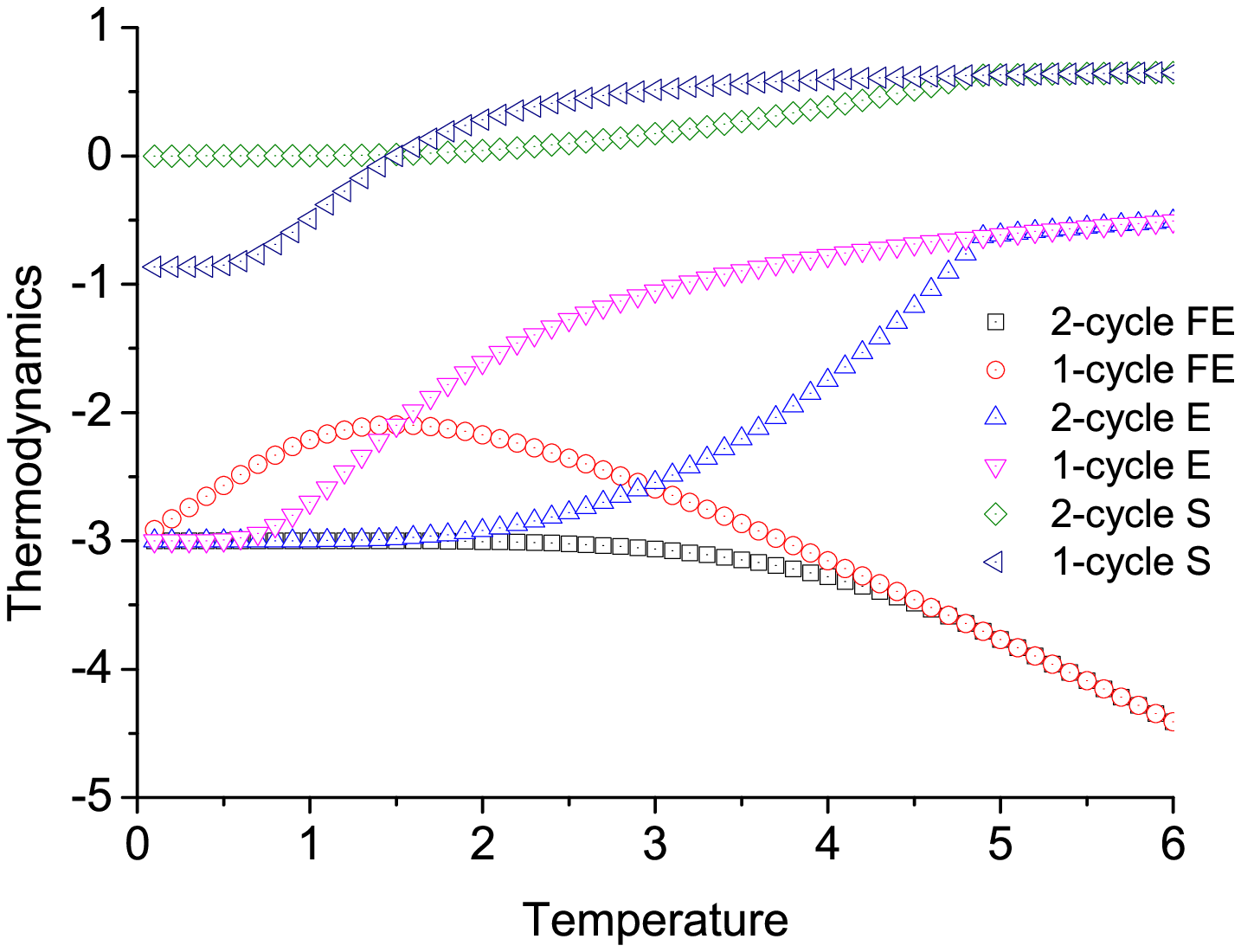}%
\caption{The reference thermal properties of the 2-cycle and 1-cycle case for
$r=3$ HL ($J=-1$ and other parameters are zero).}%
\label{fig8}%
\end{center}
\end{figure}

\begin{figure}
[ptb]
\begin{center}
\includegraphics[width=0.8\textwidth]{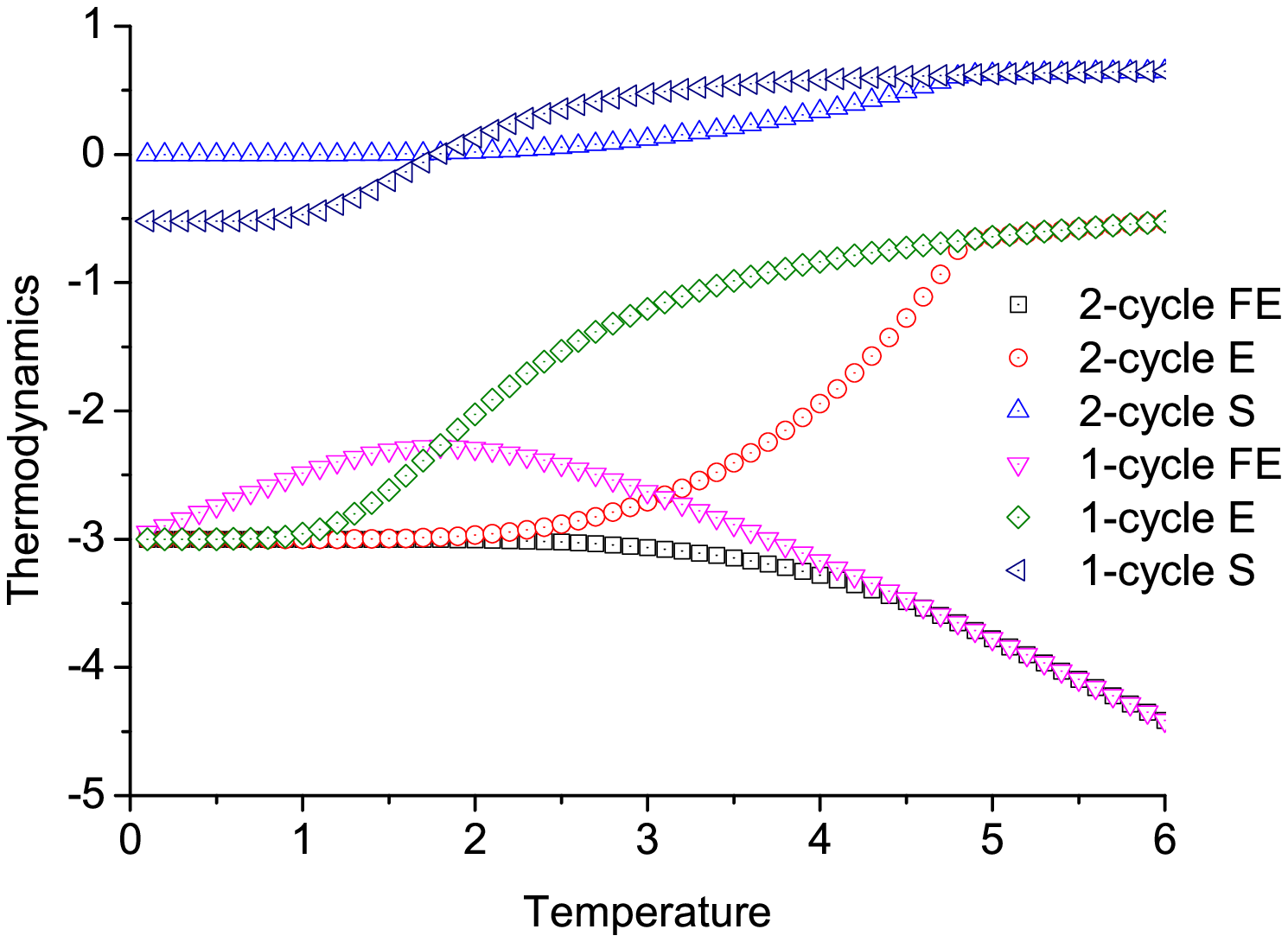}%
\caption{The reference thermal properties of the 2-cycle and 1-cycle case for
CRL ($J=-1$ and other parameters are zero).}
\label{fig9}%
\end{center}
\end{figure}

Figure \ref{fig8} and \ref{fig9} show the thermodynamics of $r=3$ HL and CRL with reference case. Basically, like the behaviors of solutions, the two models exhibit similar thermal behaviors with slight difference. Herein, we will discuss the general transitions occurring on each type of solution, and re-examine the properties of models those have been concluded from the solutions. 

For the 2-cycle solutions, there is merely the melting process that can be observed. At the melting transition, the entropy of 2-cycle solution falls dramatically
due to the crystallization, The rapid drop in the entropy of the 2-cycle reflects that the simple
Ising model, which we will call the reference model in this work, is
not realistic enough to give a discontinuous melting. Although a continuous
transition seems contradicting to the general idea of \textquotedblleft
melting process\textquotedblright as a first-order transition, the phenomenon of transition from
the ordered to the disordered state should still be treated as the melting
process. Further studies also confirm that particular parameters setup can produce a discontinuous transition for the 2-cycle case, this part will not be detailed in this paper.

On the other hand, following the behavior of the supercooled liquid, with the cooling process the entropy of 1-cycle solution continues to decrease gradually at and below $T_{\text{M}}$, and there occurs no transition. At some point, the entropy of the supercooled liquid decreases faster than the crystal, and rapidly goes below the crystal entropy then to negative values, which demonstrates the Kauzmann paradox \cite{Kauzmann,Debenedetti}, i.e. the ideal glass transition. 

The Kauzmann paradox was originally referred to where the entropy of supercooled liquid becomes lower than the crystal entropy. This is because that, at the same temperature the kinetic (vibration) entropy of
crystal and supercooled liquid are the same, and the entropy of crystal was
approximated to be the kinetic entropy, then the difference between the
entropy of crystal and supercooled liquid, i.e. the excess entropy, equals to
the configurational entropy, which cannot be negative. However, the entropy of
crystal at non-zero temperature is not only contributed by the kinetic entropy, it
also has entropy from defects. In this way it is possible to have a
supercooled liquid state with less defects and lower entropy than the crystal
state at the same temperature. Therefore, we will simply take the negative entropy of the supercooled liquid as the paradox, and the temperature where negative $S$ occurs is defined as the Kauzmann temperature $T_{\text{K}}$. This clarification is important, since in later section we will show a special case, in which we still have negative excess entropy but the negative entropy itself vanishes.

\subsection{The effect of $J_{p}$}

\begin{figure}
[ptb]
\begin{center}
\includegraphics[width=0.8\textwidth]{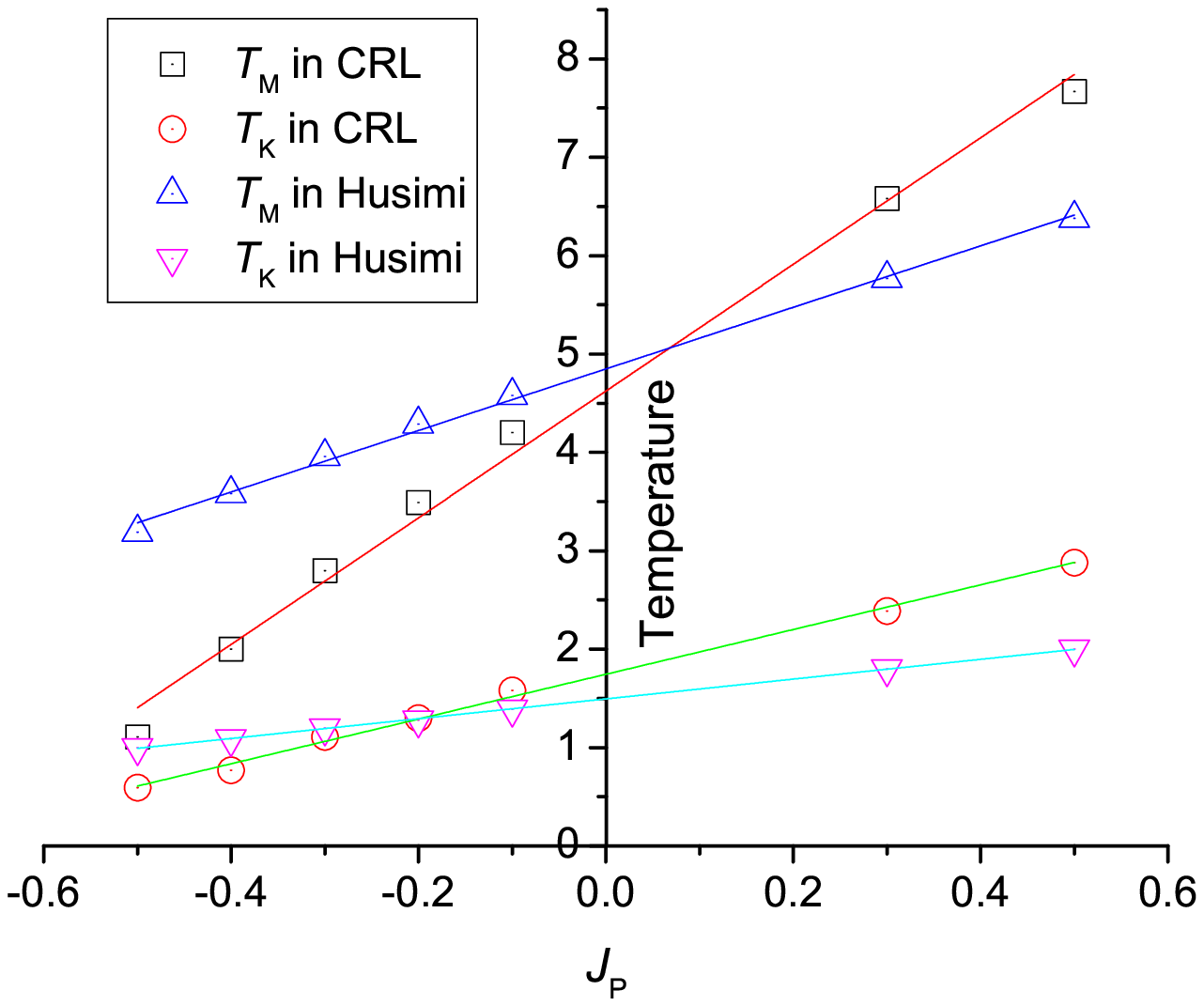}%
\caption{The transition temperature variation with different $J_{P}$ in CRL
and the HL of $r=3$.}%
\label{fig10}%
\end{center}
\end{figure}

Beside the interaction $J$ between the nearest spins, the interactions between
spins farther apart also play important roles in our lattice model. With the
set of various values of these parameters, the model is capable to simulate
many different cases, i.e. to manipulate the transition temperatures. As we
modify the interaction along the surface diagonal $J_{P}$, it significantly
changes the transition temperatures as shown in Figs. \ref{fig10}.

In both lattices, the increment of $J_{P}$ enhances the transition $T$ in a linear
relationship. From Eq. \ref{e_alpha} it is easy to understand that the negative value of
$J_{P}$ competes with negative $J$. The negative $J$ prefers different states
of the nearest spins and consequently the same states of the spins on the surface
diagonal, while negative $J_{P}$ forces the diagonal spins to be in different
states. Thus, in both the CRL and HL the decrease of $J_{P}$
lowers the transition temperature, that is, makes the system unstable and easier to undergo a transition. On the other hand, positive $J_{P}$ favors antiferromagnetic
spin configuration and increases the transition temperature for the same
reason. In CRL, the role of $J_{P}$ is more significant than it is in the
HL, because a cubic unit has 12 surface diagonal interactions and
12 nearest neighbor interactions, while a square unit has 4 $J$'s and 2 $J_{P}$'s
therefore the diagonal interactions have less weight.

\subsection{The effect of $J^{\prime}$}

\begin{figure}
[ptb]
\begin{center}
\includegraphics[width=0.8\textwidth]{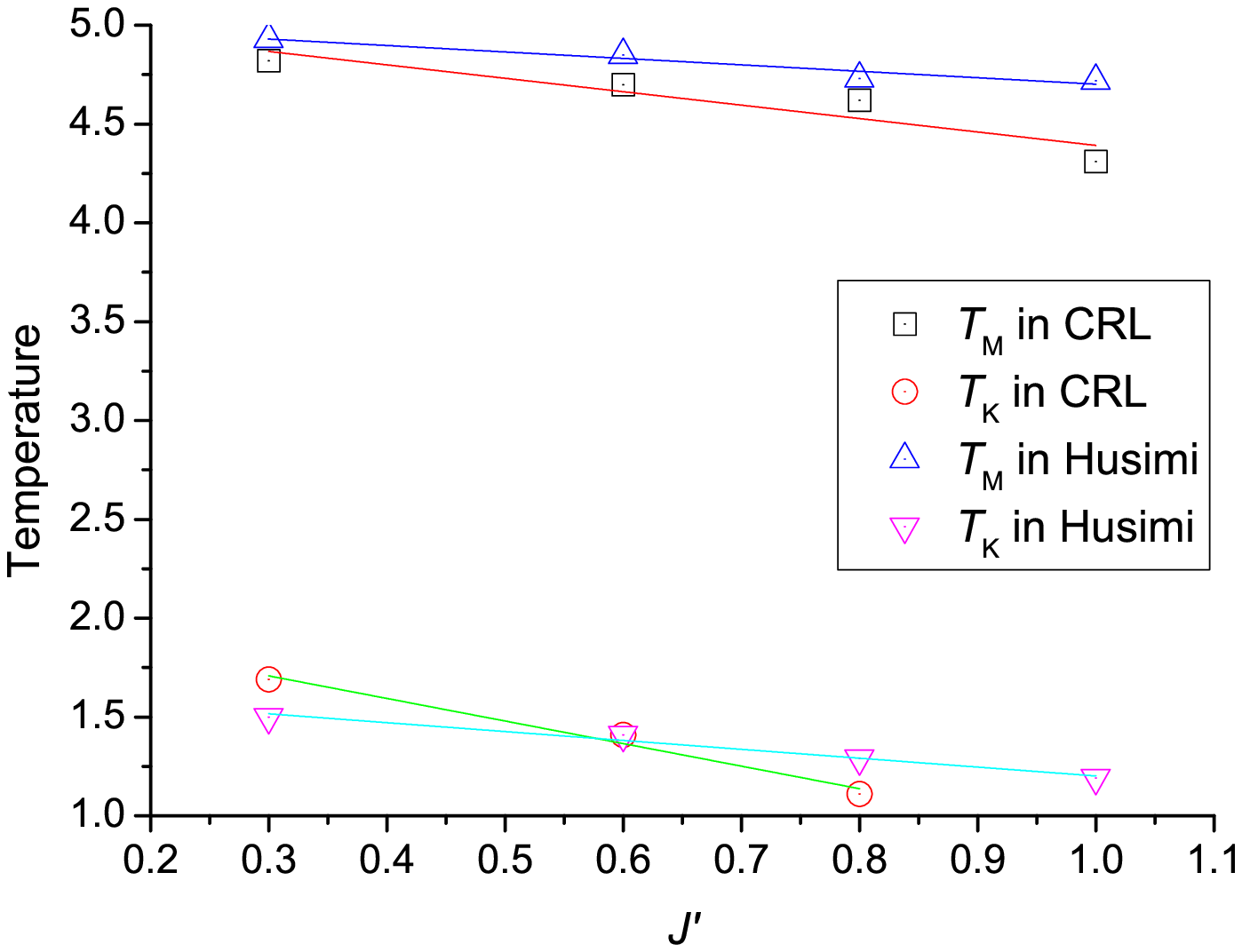}%
\caption{The transition temperature variation with different $J^{\prime}$ in
CRL and the HL of $r=3$.}%
\label{fig11}%
\end{center}
\end{figure}

An important feature of $J^{\prime}$ is that its sign (positive or negative)
does not affect the solutions. With the increase of $J^{\prime}$, the
transition temperatures decrease by a small slope in the CRL, and
even smaller in the HL. That is because the three-spins
interactions would disturb the ordered state formation; nevertheless its
contribution is very small in the total energy in either cases (Fig.
\ref{fig11}).

\begin{figure}
[ptb]
\begin{center}
\subfloat[]{
\includegraphics[width=0.8\textwidth]{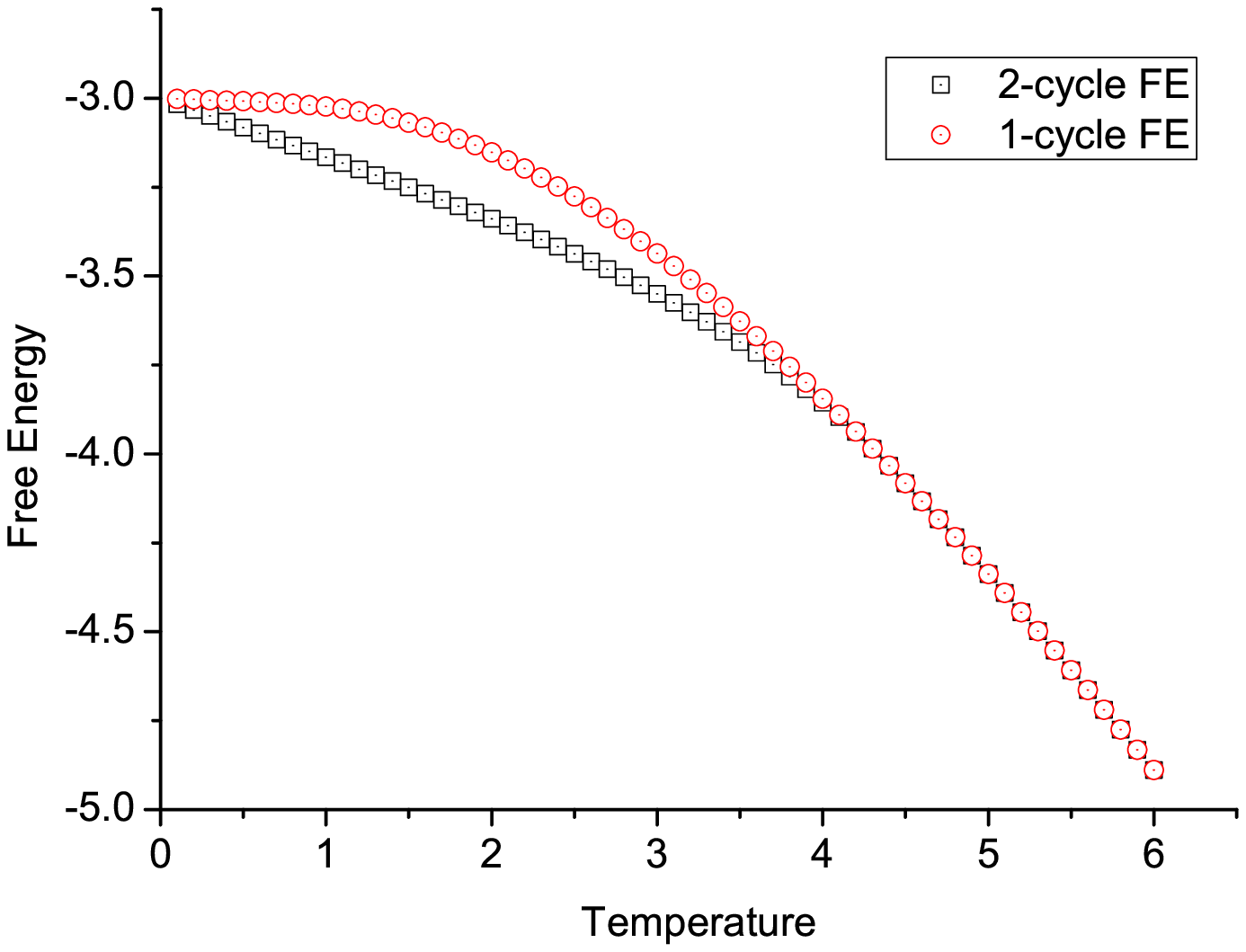}
}\\
\subfloat[]{
\includegraphics[width=0.8\textwidth]{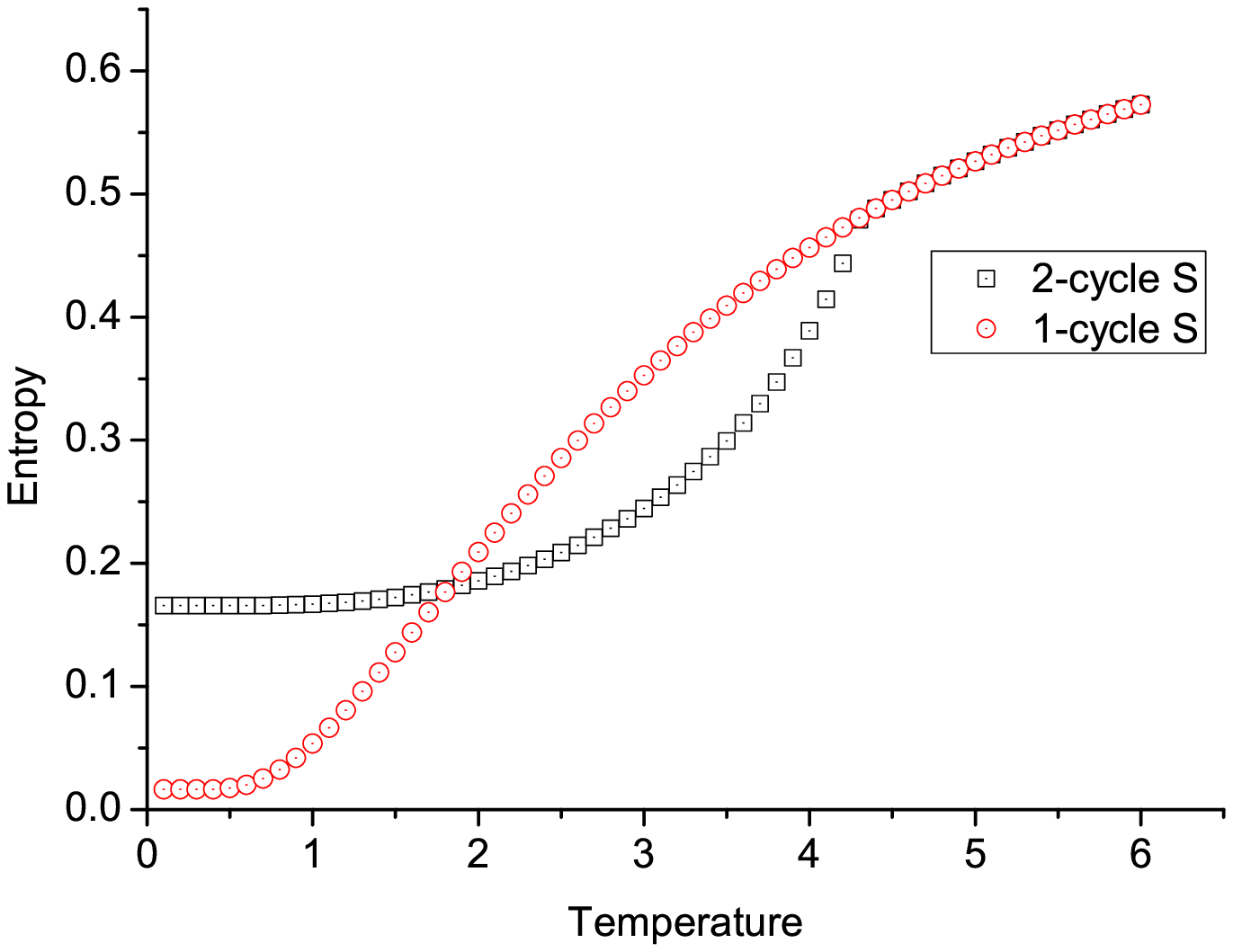}
}
\caption{The (a) free energy and (b) entropy of the extraordinary case of CRL model with $J^{\prime}=1$. The Kauzmann paradox vanishes for the 1-cycle solution.}
\label{fig12}%
\end{center}
\end{figure}

Another interesting effect of $J^{\prime}$ is that, although the transition
temperature changes weakly with $J^{\prime}$, it changes the thermodynamic
curves dramatically, and with $J^{\prime}=1$ it could even destroy the ideal
glass transition by having the supercooled liquid entropy positive till zero
temperature, as shown in Fig. \ref{fig12}. In this situation, we still have the original Kauzmann paradox without permitting the lower entropy of supercooled liquid than the crystal, nevertheless by our definition the negative entropy vanishes and the paradox does not hold anymore.

\subsection{The effect of $J^{\prime\prime}$}

\begin{figure}
[ptb]
\begin{center}
\includegraphics[width=0.8\textwidth]{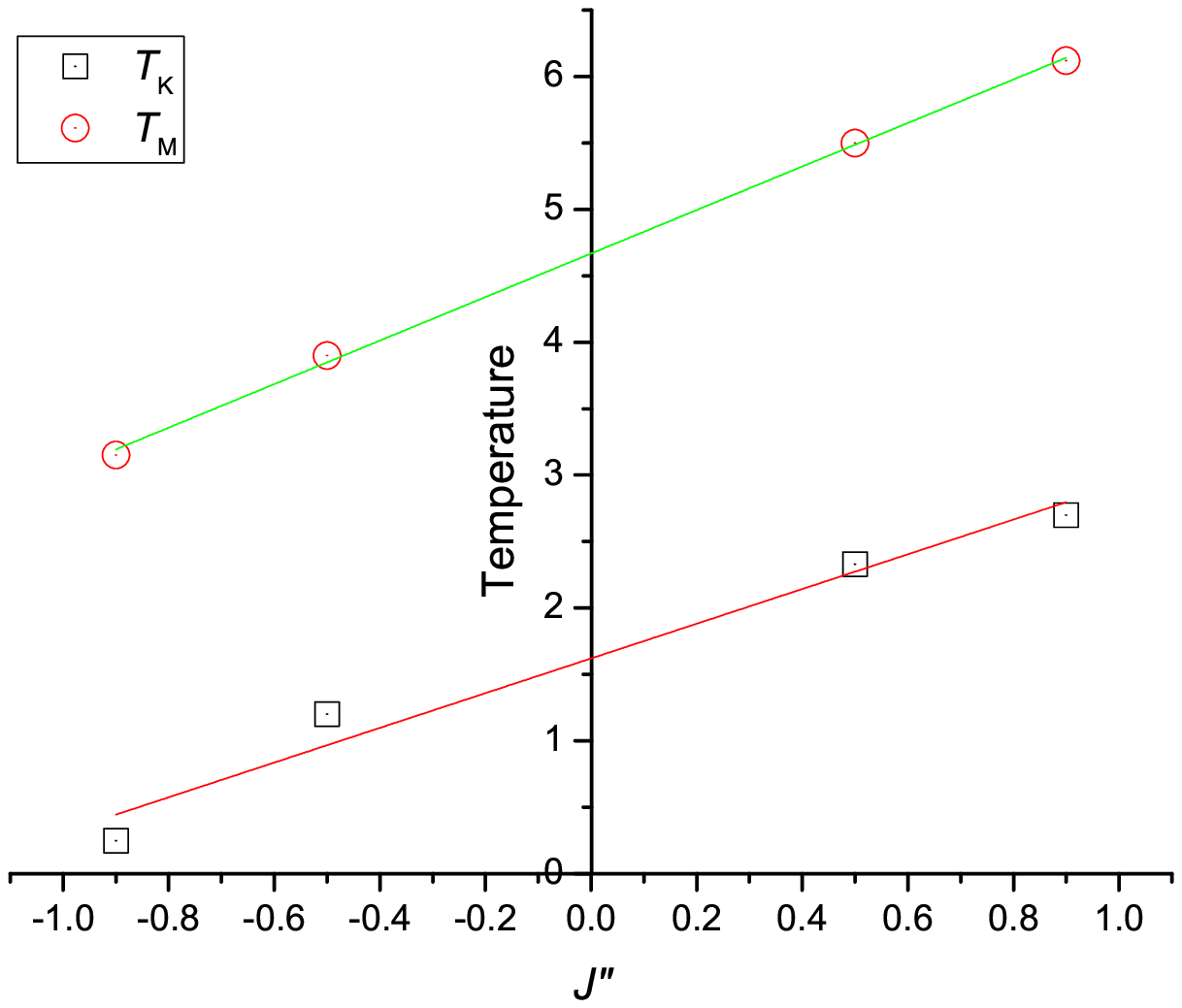}%
\caption{The transition temperature variation with different $J^{\prime\prime
}$ in CRL.}
\label{fig13}
\end{center}
\end{figure}

The roles of four-spins interaction $J^{\prime\prime}$ in CRL and HL are very different. There are six squares in one cubic unit in the CRL. The value of $J^{\prime\prime}$ affects the spins states in the same way as $J_{P}$, i.e. it increases the transition temperature. However it does not
change $T_{\text{M}}$ and $T_{\text{K}}$ in the HL because the
4-spins interaction is mainly the interaction of the whole unit, thus its effect
is like a magnetic field $H$ for each single spin (Fig. \ref{fig13}).

\section{Conclusion}

The Ising model on a recursive lattice enables us to do exact calculations
without approximation to study the supercooled liquid, the crystal
formation and phase transitions. We have constructed and compared a multi-branched 2-D
HL and a cubic recursive lattice of the same coordination number $q$. The
branch number is set to be 3 in the multi-branched 2-D HL to make
$q$ to be 6. Because of the identical $q$, both
lattice are designed to approximate the regular 3-D lattice. We applied the
antiferromagnetic Ising model on the two lattices, in which the alternative spin
arrangement represents the ordered phase (crystal state), while the
homogeneous spins arrangement represents the metastable state, i.e. the
liquid and the supercooled liquid states. 

A set of solutions relating to the
probability that one site being occupied by a particular spin state can be
exactly calculated on the lattice from the ratio of partial partition
functions. Two types of solutions, representing the ordered and disordered states
respectively, give the thermodynamics of crystal and metastable states. The
behavior of the free energy and the entropy of two states allows us to locate
the melting transition at $T_{\text{M}}$ and the Kauzmann temperature
$T_{\text{K}}$. The $T_{\text{M}}$ is determined by the differing of two
states with temperature decreasing (cooling process), above $T_{\text{M}}$ the two solutions have
the same disordered behavior as liquid. When the system goes below 
$T_{\text{M}}$, an ordered state (from 2-cycle solution) appears with lower
free energy and entropy, while the disordered solution (1-cycle)
still exists and the system is possible to continue in supercooled state
without phase transition. The entropy of this disordered state gradually
decreases to zero and then becomes negative at a non-zero temperature, which is
determined to be the ideal glass transition temperature, or Kauzmann
temperature $T_{\text{K}}$.

The transition temperatures determined on both lattices are very close to each
other, and fairly agree with the results from other methods such as MC
simulation and series analysis. The results confirm our expectation that: (1) The 
CRL is a good approximation of regular 3-D cubic lattice; (2) With the same
coordination number the multi-branched 2-D HL is also capable to
describe systems in higher dimension. It is also concluded that the 3-branched Husimi
lattice provides a less versatile and accurate method compared to the CRL. For
example, the CRL provides a transition temperature closer to that obtained
by series analysis than the one obtained by the HL. However, the
multi-branched HL calculation is simpler and less time-consuming,
which still holds merits in applications.

In addition to the nearest neighbor interactions, the effects of second
nearest interaction, triplet and quadruplet interaction are also
studied in this work. All their effects are similar for the two lattices
except $J^{\prime\prime}$. This enables us to obtain expected transition
temperature or thermodynamic behaviors by adjusting the interaction parameters
for an extensive use of our models.


\begin{thebibliography}{99}                                                                         

\bibitem {Kauzmann}W. Kauzmann, Chemical Reviews \textbf{43}, 2 (1948).

\bibitem {glass1}\textit{The Class Transition and the Nature of the Glassy State},
edited by M. Goldstein and R. Simha, Ann. N. Y. Acad. Sci. 279 (1976).

\bibitem {glass2}W. Gotze, \textit{Liquids, Freezing and the Glass Transition}, edited by J.
P. Haasma, D. Levesque, and J. Zinn-Justin, North-Holland, Amsterdam (1991).

\bibitem {glass3}P. G. Debenedetti, \textit{Metastable Liquids: Concepts and
Principles}, Princeton University Press, Princeton, New Jersey (1996).

\bibitem {glass4}M. M\'{e}zard and G. Parisi, J. Phys.: Condense. Matter. \textbf{12}
6655 (2000).

\bibitem {Debenedetti}P. G. Debenedetti and F. H. Stillinger, Nature \textbf{410},
259 (2001).

\bibitem {Bethe}H. A. Bethe, Proc. Roy. Soc. London Ser A \textbf{150}, 552 (1935).

\bibitem {Husimi} K. Husimi, J. Chem. Phys. \textbf{18}, 682 (1950).

\bibitem {pdg_prl_reliable}P. D. Gujrati,  Phys. Rev. Lett. \textbf{74}, 809 (1995). 

\bibitem {PDG2} J. Ryu and P. D. Gujrati, J. Chem. Phys. \textbf{107}, 1259 (1997). 
\bibitem {PDG3} P. D. Gujrati, J. Chem. Phys. \textbf{108}, 5089 (1998). 

\bibitem {PDG4}F. Semerianov and P. D. Gujrati, Phys. Rev. E \textbf{72}, 011102 (2005).

\bibitem {exact} R. J. Baxter, \textit{Exactly Solved Models in Statistical Mechanics}, Academic Press, London (1982) 1st ed., p. 47.

\bibitem {12}A. J. Banchio and P. Serra, Phys. Rev. E \textbf{51}, 2213 (1995).

\bibitem {13}R. A. Zara and M. Pretti, J. Chem. Phys. \textbf{127}, 184902 (2007).

\bibitem{unit1} J. A. Verges and F. Yndurain, J. Phys. F: Met. Phys. \textbf{8}, 873 (1978).

\bibitem{Geertsma} W. Geertsma and J. Dijkstra, J. Phys. C: Solid State Phys. \textbf{18}, 5987 (1985). 
\bibitem{unit2} J. F. Stilck and M. J. de Oliveira, Phys. Rev. A \textbf{42}, 5955 (1990).

\bibitem {Ran1} R. Huang and C. Chen, Commun. Theor. Phys. \textbf{62}, 749 (2014).

\bibitem {Ising}E. Ising, Z. Phys. \textbf{31}, 253 (1925).

\bibitem {Onsager}L. Onsager, Phys. Rev. \textbf{65}, 117 (1944).

\bibitem {Ohzeki} M. Ohzeki, Phys. Rev. E \textbf{87}, 012137 (2013).

\bibitem {Wu}B. M. McCoy and T. T. Wu, \textit{The two-dimensional Ising
model}, Harvard Univ. Press, Cambridge, Massachusetts (1973).

\bibitem {cavity} M. M\'{e}zard and G. Parisi, Eur. Phys. J. B \textbf{20}, 217 (2001).

\bibitem {23}A. P. Ramirez, Annu. Rev. Mater. Sci. \textbf{24}, 453 (1994).

\bibitem {24}L. D. Landau and E. M. Lifshitz, \textit{Statistical Physics}, 3rd
edition, Part I, Pergamon Press, Oxford (1986).

\bibitem {25}D. Ruelle, Physics (Utrecht) \textbf{113A}, 619 (1982).

\bibitem {26}A. M. Ferrenberg and D. P. Landau, Phys. Rev. B \textbf{44}, 5081 (1991).

\bibitem {27}E. E. Reinerhr and W. Figueiredo, Phys. Rev. \textbf{160}, 393 (1967).

\bibitem {28}Z. Racz, Phys. Rev. B \textbf{21}, 4012 (1980).

\end{thebibliography}
\end{document}